\documentclass[prd,twocolumn,superscriptaddress,showpacs,nofootinbib,preprintnumbers]{revtex4}

\usepackage{amsmath}
\usepackage{amsfonts}
\usepackage{graphicx}
\usepackage{dcolumn}
\usepackage{hyperref}

\def\bs{\begin{subequations}}
\def\es{\end{subequations}}

\newcommand{\beq}{\begin{equation}}
\newcommand{\eeq}{\end{equation}}
\newcommand{\beqa}{\begin{eqnarray}}
\newcommand{\eeqa}{\end{eqnarray}}

\newcommand{\calr}{{\cal R}}

\newcommand{\la}{\langle}
\newcommand{\ra}{\rangle}

\begin{document}

\title{Testing for double inflation with WMAP}

\author{David Parkinson}
\affiliation{Institute of Cosmology and Gravitation, University of 
Portsmouth,
Mercantile House, Portsmouth PO1 2EG, UK}
\author{Shinji Tsujikawa} 
\affiliation{Department of Physics, Gunma National College of 
Technology, Gunma 371-8530, Japan}
\author{Bruce A. Bassett}
\affiliation{Department of Physics, Kyoto University, Kyoto, Japan}
\affiliation{Institute of Cosmology and Gravitation, University of 
Portsmouth,
Mercantile House, Portsmouth PO1 2EG, UK}
\author{Luca Amendola}
\affiliation{INAF/Ossevatorio Astronomico di Roma, Viale, Frascati 33, 
00040 Monte Porzio Catone (Roma), Italy}

\date{\today{}}
\begin{abstract}
With the WMAP data we can now begin to test realistic models 
of inflation involving multiple scalar fields. These  naturally lead to
correlated adiabatic and isocurvature (entropy) perturbations with a running 
spectral index.  
We present the first full (9 parameter) likelihood
analysis of double inflation with WMAP data and find that despite the 
extra freedom, supersymmetric hybrid potentials are strongly 
constrained with less than $7\%$ correlated isocurvature component allowed 
when standard priors are imposed on the cosomological parameters. 
As a result we also find 
that Akaike \& Bayesian model selection criteria rather strongly 
prefer single-field inflation, just as equivalent analysis 
prefers a cosmological constant over dynamical 
dark energy in the late universe. 
It appears that simplicity is the best guide to our universe.                
\end{abstract}

\maketitle

\section{Introduction}

Our universe shows evidence of complexity and, at the same time, great 
simplicity. 
Our universe appears entirely consistent with being a ``double-de Sitter 
sandwich" - radiation and
 matter dominated phases caught between two de Sitter phases at low and 
high energies respectively.  

Recent work \cite{jochen,params} has shown that a cosmological constant 
provides a better fit 
than dynamical dark energy to current CMB and SNIa data if one computes 
the Bayesian evidence or uses information criteria for model selection. 
In this paper we will show that, at least within a class of double, hybrid 
inflation models, 
the same is true for the early universe. 
One might envisage various infrared-ultraviolet dualities 
to explain such 
behaviour.

Despite this apparent ``asymptotic blandness " there is interesting tentative 
evidence to the contrary. 
The WMAP data show unusual characteristics such as 
``oscillations" \cite{oscil} which may disappear with more data 
or may be the first signs of new physics. 
Similarly there is evidence for a feature in the power spectrum 
\cite{feature} which can easily be produced by the subtle dynamics of 
multiple light scalar fields during inflation.

Multiple light fields during 
inflation automatically widens the narrow 
predictions of single-field 
inflation for now there are multiple entropy 
perturbations \cite{PS}-\cite{GV}  
which are, in general, correlated to some degree with the standard adiabatic 
mode \cite{Langlois}-\cite{Crotty}. 
Correlations are produced when the valley of the effective potential is 
curved \cite{Gordon} and this also
leads to non-gaussianity \cite{nongauss}.  
Since the effective masses of the various fields 
typically depend on the vacuum 
expectation values of the other (dynamical) fields these are 
time-dependent and 
can cause violations of standard slow-roll conditions and spectral indices 
for the perturbations which run with scale \cite{Tsuji}.

This is a crucial aspect of this present work because previous analyses 
of correlated adiabatic and entropy (isocurvature) perturbations have 
always assumed power-law spectra for all the perturbations 
\cite{Amen,BMT,WM,VM,Crotty}. 
When applied to the WMAP data they found that with standard priors on 
cosmological parameters the degree of correlation allowed is 
small (although see \cite{cons}). Allowing running of the spectral index,
at least in the supersymmetric hybrid models we study, does not change 
this conclusion.

The code we have developed allows us to 
numerically study 
any inflationary model without approximation 
(except in the treatment of spinodal/tachyonic instabilities) 
and builds on that used in \cite{Tsuji}. 
Future work will consider more fields where sharp features can occur, 
something which does not occur in the two-field double-inflation models we 
study here. 

\section{General formalism}

We consider two minimally coupled scalar fields, $\phi$ and $\chi$,
with an effective potential $V(\phi,\chi)$. 
Our main interest is the case of double inflation 
in which two stages of inflation are realized.
General scalar metric perturbations  about
the flat Friemann-Lemaitre-Robertson-Walker background can be 
written as (see e.g. \cite{Gordon})
%%%%%%%%%%
\begin{eqnarray}
ds^{2}&=&-(1+2A)dt^{2}+2a(t)B_{,i}dx^i dt \\ \nonumber
& &+a^{2}(t)[(1-2\psi)\delta _{ij}+2E_{,ij}]dx^{i}dx^{j}\,,
\label{metric}
\end{eqnarray}
%%%%%%%%%%
where $a(t)$ is the scale factor. 
The comoving curvature perturbation in our two-field system is 
then given by 
%%%%%%%%%%
\begin{eqnarray}
\calr \equiv \psi +\frac{H(\dot{\phi }\delta \phi +
\dot{\chi }\delta \chi )}{\dot{\phi }^{2}+\dot{\chi }^{2}}\,,
\label{calR}
\end{eqnarray}
%%%%%%%%%%
where $H \equiv \dot{a}/a $ is a Hubble rate, and $\delta \phi$ 
and $\delta \chi $ are the perturbations of the fields 
$\phi $ and $\chi $, respectively.

The perturbation equations
are given in Refs.~\cite{Gordon,Bartolo}
and one can numerically evaluate the power spectrum,
${\mathcal{P}}_{\calr } \equiv (k^3/2\pi^2)|\calr|^2$,
at the end of inflation \cite{Tsuji} 
(here $k$ is comoving momentum).
In the multi-field system we also need to account for the
spectra of isocurvature perturbations, ${\mathcal{P}}_{S}$, and 
correlated adiabatic and isocurvature perturbations, 
${\mathcal{P}}_{C}$ (see Ref.~\cite{Bartolo,Tsuji} 
for their definitions).
The quantity $r_C$ defined by
$r_{C}={\mathcal{P}}_{C}/\sqrt{{\mathcal{P}}_{\mathcal{R}}
{\mathcal{P}}_{S}}$
is the measure of the strength between adiabatic and isocurvature 
perturbations.

The system possesses several model parameters
associated with the potential.
We assume slow-roll conditions apply 
$|\ddot{\phi}| \ll |3H\dot{\phi}|$ 
and $|\ddot{\chi}| \ll |3H\dot{\chi}|$, 
for the initial conditions of 
background fields, so that
the $\ddot{\phi}$ and $\ddot{\chi}$ terms are neglected.
Then the initial conditions of $\dot\phi$ and $\dot\chi$ are
determined by $\phi_{\rm in}$ and $\chi_{\rm in}$ 
(the subscript ``in'' denotes the initial values).
We perform the likelihood analysis over the initial 
conditions $\phi_{\rm in}$, $\chi_{\rm in}$.

Note that the number of inflationary model parameters depends explicitly 
on the inflaton potential 
and typically requires at least three parameters 
in the context of double inflation.

We impose the condition that the total number of $e$-folds
during inflation must exceed
$N_{\rm T}=50$ to solve flatness and horizon problems.
We find the cosmologically relevant perturbation modes
with comoving wavenumbers $k$ and numerically evolve the background 
and all perturbation equations through inflation, giving us 
the three power spectra
$({\mathcal{P}}_{\calr}, {\mathcal{P}}_S, {\mathcal{P}}_C)$, as described
in \cite{Tsuji}.

It is important to solve the perturbation equations without approximation
right up to the end of 
inflation, since the curvature perturbation is not necessarily conserved
after Hubble radius crossing \cite{GW}, 
unlike the case of single-field inflation.

The resulting data: $P_{i}=\log {\mathcal{P}}(k_{i})$
 given at
a wave number of $x_{i}=\log k_{i}$, $i=1,..m$, are 
optimally fitted with a polynomial 
function $P_{\rm fit}(x)=a+bx+cx^{2}$
by minimizing 
$\bar{\chi}^{2}=\sum _{i}(a+bx_{i}+cx_{i}^{2}-P_{i})^{2}$. 

For each set of parameters we 
derive the best-fit coefficients $a, b, c$ for each of the three power spectra. 
It is worth mentioning 
that the coefficients $b$ and $c$ are intimately linked to 
the spectral index $n_s$ and its running of scalar perturbations by the 
relations $n_s=b+1$ and $\alpha_s=2c$. 
We check that our fitting method
agrees very well with numerically obtained power spectra 
and is sufficient to accurately capture any running
of the spectral index over cosmologically relevant 
scales.

We assume, as is standard, that the 
field $\phi$ decays to ordinary matter
like photons, neutrinos and baryons,
whereas the field $\chi$ decays into cold 
dark matter  (CDM) \cite{Langlois,Gordon}.
In this case the mixing between two scalar fields is negligible
and the CDM isocurvature perturbations and correlations remain after reheating. Relaxing this assumption
will introduce extra parameters into the analysis. 

The CMB temperature anisotropies are given in general by
%%%%%%%%%
\begin{eqnarray}
C_{\ell}=(4\pi )^{2}\int k^{2}dk\Delta _{\ell }^{2}
\la k,\tau _{0} \ra,  
\end{eqnarray}
%%%%%%%%%
where $\Delta _{\ell }(k,\tau _{0})$ is the $\ell $-multipole of
the $k$-th wavenumber temperature anisotropy 
at the present time $\tau_0$. 
For a general set of correlated initial conditions, one has
%%%%%%%%%
\begin{eqnarray}
\Delta (k,\tau _{0})=P_{\mathcal{R}}^{1/2}
\Delta _{\mathcal{R}}(k,\tau _{0})
+P_{S}^{1/2}\Delta _{S}(k,\tau _{0}),
\end{eqnarray}
%%%%%%%%%
where 
$ \la \Delta _{\mathcal{R}}^{2}(k,\tau _{\rm in}) \ra=
\la\Delta _{S}^{2}(k,\tau _{\rm in})\ra=1$ and 
$\la\Delta _{\mathcal{R}}(k,\tau _{\rm in})
\Delta _{S}(k,\tau _{\rm in}) \ra=r_C$. 
Then we get 
%%%%%%%%%
\begin{eqnarray}
C_{\ell } &=& (4\pi )^{2}\int k^{2}dk[P_{\mathcal{R}}
\Delta _{\ell ,\mathcal{R}}^{2}+P_{S}\Delta _{\ell ,S}^{2}+2r_{c} \la
\Delta _{\ell ,\mathcal{R}}\Delta _{\ell ,S} \ra] \nonumber \\
& \equiv &C_{\ell,\mathcal{R}}+C_{\ell ,S}+2C_{\ell ,C}\,.
\end{eqnarray}
%%%%%%%%%
It is possible to obtain the three multipole spectra required for
any general set of initial perturbations using the following
simple scheme. Let us denote $C(A_{1},A_{2})$ as
the $C_{\ell}$ spectrum
obtained with completely correlated initial conditions with a given
adiabatic spectrum $A_{1}$ and given 
isocurvature spectrum $A_{2}$.
A typical Boltzmann code can produce only $C(A_{1},0)$ (pure adiabatic),
$C(0,A_{2})$ (pure isocurvature) or $C(A_{1},A_{1})$ (completely
correlated mixture of adiabatic and isocurvature with the same initial
spectrum). It is not difficult to see that the general spectrum is given by:
%%%%%%%%%
\begin{eqnarray}
C(A_{1},A_{2}) & = & 
C(A_{1},0)+C(0,A_{2})+2[C(A_{12},A_{12}) \nonumber \\
& &-C(0,A_{12})-C(A_{12},0)]\,,
\end{eqnarray}
%%%%%%%%%
where in our case $A_{12}=\sqrt{P_{C}}$ , $A_{1}=\sqrt{P_{\mathcal{R}}}$
and $A_{2}=\sqrt{P_{S}}$. One needs therefore five evaluations for
each combination of $P_{\mathcal{R},S,C}$.  
We make use of a modified version of 
the CAMB Boltzmann solver \cite{Lewis} to evaluate the CMB power spectrum 
by this scheme.  

In addition to $\phi_{\rm in}$, $\chi_{\rm in}$ and the 
inflationary potential parameters discussed in the
next section, we varied 4 cosmological
parameters: $\Omega_b h^2$, $\Omega_c h^2$, $\tau$, $H_0$; 
namely the baryon and cold dark matter density, 
the  reionisation optial depth and the Hubble 
constant today.  We assume spatial flatness, so 
$\Omega_{\Lambda} = 1 - \Omega_b - \Omega_c$.

It is well-known that the allowed ranges for
these parameters has a large impact on the acceptable
amount of correlated isocurvature perturbations 
\cite{cons2}. We choose fairly standard priors, 
allowing the above variables to vary in the ranges:
$\tau \in [0,0.3]$, $H_0 \in [50,90]$ with
$\Omega_b$ and $\Omega_c$ both varying over the full
unit interval, $[0,1]$. We choose
very wide domains for $\phi_{\rm in}$ and $\chi_{\rm in}$ and 
found that the results depended very weakly on 
these boundaries. 

We then use the first year WMAP TT and TE 
data \cite{Verde} in our analysis to constrain the 
various parameters.

%%%%%%%%%%%%%%%%%%%%%%%%%%%%%%%%
\section{A realistic double inflation model and likelihood results}
%%%%%%%%%%%%%%%%%%%%%%%%%%%%%%%%

Let us consider a fairly realistic 
multi-field inflation model with potential
\begin{eqnarray}
V=\frac{\lambda }{4}\left(\chi ^{2}-\frac{M^{2}}{\lambda }\right)^{2}+
\frac{1}{2}g^{2}\phi ^{2}\chi ^{2}+\frac{1}{2}m^{2}\phi ^{2}\,,
&  &
\label{po}
\end{eqnarray}
corresponding to the original version of the hybrid inflation
\cite{Linde}. This is closely linked with those obtained in supersymmetric
theories \cite{Copeland,Randall,LR}, 
which generically leads to a very strong correlation
between the adiabatic and isocurvature perturbations due to the presence of
a tachyonic instability between the two phases 
of inflation \cite{Tsuji}. In this work we concentrate on the
supersymmetric case with $g^{2}/\lambda =2$.
Then we have three potential parameters: $\lambda$, $M$ and $m$, 
which are constrained by our likelihood analysis.

We can have two stages of inflation for the potential (\ref{po})
depending upon the model parameters.
One corresponds to the stage with $\phi>\phi_c \equiv M/g$ driven by
the slow-roll evolution of $\phi$
during which the potential is approximately described
by $V\simeq M^{4}/4\lambda +m^{2}\phi ^{2}/2$.
Another inflationary stage is the one with $\phi<\phi_c$ driven
by the field $\chi$ with a tachyonic instability.

When the condition $M \gtrsim m$ is satisfied, then
$M^{4}/\lambda \gtrsim m^{2}\phi_c^2$,
and so the Hubble rate is roughly constant with a value
$H=\sqrt{2\pi/3\lambda}M^2/m_p$,
around $\phi=\phi_c$ (here $m_p$ is the Planck mass). 
We can estimate the condition for double
inflation by estimating the effective masses of the
two fields, i.e., $m_\phi^2 \simeq m^2$ and $m_\chi^2 \simeq 
g^2\phi^2-M^2$.
Double inflation occurs when both of the masses of 
the two fields are smaller than 
$H$, which gives the condition
\begin{eqnarray}
\label{double1}
& &M^2 \gtrsim mm_p \sqrt{3\lambda/2\pi}\,, \\
& &(M/m_p)^2 \gtrsim 3\lambda/2\pi  \,.
\label{double2}
\label{con}
\end{eqnarray}

We are mainly interested in the double inflation scenario in which
the second stage of inflation occurs after 
the symmetry breaking.
Since $m_\chi$ is smaller than $H$ around $\phi=\phi_c$,
the field $\chi$ is hardly suppressed during the
first stage of inflation, unless $\phi$ is not too much
larger than $\phi_c$.

On the other hand, when $(M/m_p)^2 \ll \lambda$, the
field $\chi$ is exponentially suppressed for 
$\phi >\phi _{c}$ and rapidly waterfalls 
toward the global minimum of the potential
after symmetry breaking.
This corresponds to the original version of
the hybrid inflation without a second stage of inflation \cite{Linde}.

In this case the homogeneous mode of $\chi$ can be 
vanishingly small relative
to its fluctuations, so the analysis using linear perturbations is not
fully trustworthy. In our work the linear perturbation equations in 
Ref.~\cite{Tsuji} are used to evaluate the
three power spectra at the end of inflation. 
While the system is stable
for the parameter range in which double inflation occurs,
we found a strong numerical instability for perturbations in the tachyonic 
instability region
when the field $\chi$ is strongly suppressed before symmetry breaking.
Thus the latter case is effectively excluded from our analysis.
In this case we need to account for the effect
of diffusion using e.g., a Fokker Planck equation \cite{Randall,Garcia},
but we do not consider this here.

In order to constrain the double inflation model given by (\ref{po}), we perform a  
likelihood analysis over 9 parameters: 5 inflationary  
$(M,\lambda,m,\phi_{\rm in},\chi_{\rm in})$ and 4 cosmological 
$(H_0,\Omega_bh^2,\Omega_ch^2,\tau)$.  

A grid-based analysis over all 9 parameters
would require a great deal of time and computing resources, and would 
still lead to very coarse sampling of the parameter space.  
Instead we conducted the analysis using a 
Markov Chain Monte Carlo (MCMC) approach.  We ran independent chains 
on different HPC facilities and used the Gelman and Rubin 
statistic to test for convergence and mixing of 
our MCMC chains, as discussed in \cite{Verde,GRS}.

Our 2d likelihood plots show two different results 
for the 1 and 2-$\sigma$ contours \cite{cosmomc}. 
The filled  contours are computed by binning 
the MCMC chains, and drawing contours around points
where the likelihood $\bar{\chi}^2$ has dropped to 0.32 (1-$\sigma$)
and 0.05 (2-$\sigma$) respectively\footnote{We use $\bar{\chi}^2$ to 
denote the standard statistical estimate of likelihood 
($-\ln {\cal L}$) in order to distinguish it from the square of the 
$\chi$ field.}.

On the other hand, the unfilled line contours show
the regions which contain $68\%$ (1-$\sigma$) and
$95\%$  (2-$\sigma$) of all the points in our chains 
(after burn-in phases are removed).
We define the burn-in point for a chain 
to be the place where the $\bar{\chi}^2$
of the chain drops below the global median 
$\bar{\chi}^2$
for the first time, as in \cite{sdss}.

%%%%%%%%%%%%%%%%%%%%%%%%%%%%%%%%
\subsection{Inflationary parameters}
%%%%%%%%%%%%%%%%%%%%%%%%%%%%%%%%

\begin{figure}[h]
\includegraphics[height=2.5in,width=3.3in]{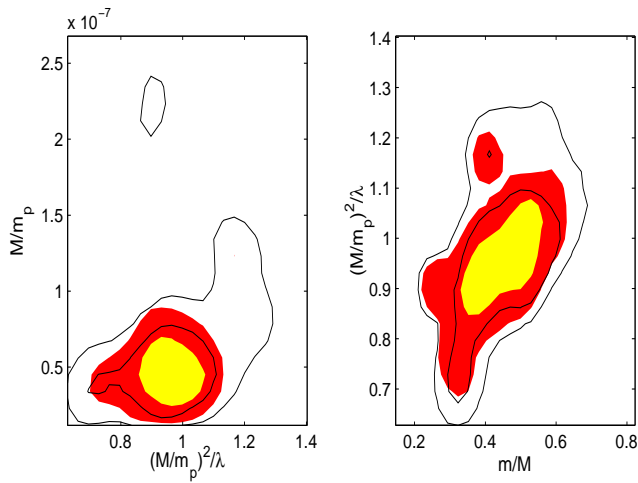}
\caption{2-dimensional likelihood constraints for the double inflationary 
parameters for the potential (\ref{po}).   
We show $1\sigma$ and 
$2\sigma$ contour bounds from $\bar{\chi}^2$ 
data (shaded contours) and the contours which contain $68\%$ and 
$95\%$ of all the points in our MC chains 
(solid lines, see discussion in the text). 
}
\label{2Dinf}
\end{figure}

In Fig.\,\ref{2Dinf} we show the 2-dimensional likelihood plots 
for various combinations of dimensionless
inflationary parameters: 
$(M/m_p, (M/m_p)^2/\lambda)$ and $(m/M, (M/m_p)^2/\lambda)$.
{}From the left panel it is clear that the 
2-$\sigma$ likelihood area is  clustered
in a small region around $\lambda \sim (M/m_p)^2$.
The square of the effective mass of $\chi$ relative to $H^2$
is given as $|m_{\chi}^2/H^2| \simeq 0.5|(\phi/\phi_c)^2-1|$ 
for $\lambda \sim (M/m_p)^2$.
Therefore $|m_\chi|$ is smaller than $H$ for 
$\phi<\phi_c$, which means
that the second stage of inflation occurs after the symmetry breaking.
When $\lambda \sim (M/m_p)^2$, the condition (\ref{double1})
translates into $M \gtrsim \sqrt{3/2\pi}\,m$.
{}From the right panel
of Fig.\,\ref{2Dinf} one finds that $m/M$ varies 
in the range $0.2<m/M<0.7$,
in which case the condition for the first stage of inflation is satisfied.
Therefore double inflation actually occurs within 
the 2-$\sigma$ likelihood region of Fig.\,\ref{2Dinf}.

%%%%%%%%%%
%%%%%%%%%%

\begin{figure}
\includegraphics[height=2.5in,width=2.5in]
{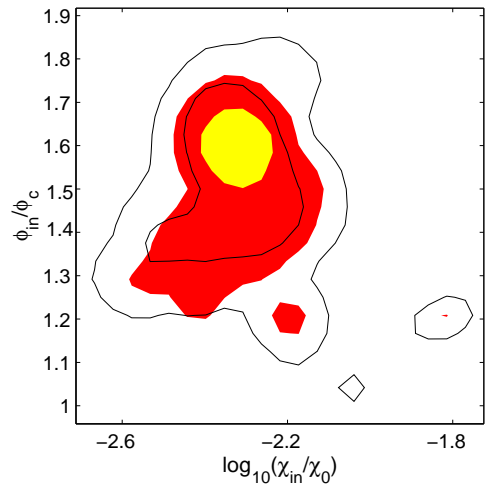}
\caption{2-dimensional likelihood constraints for the initial conditions 
of the field, $\{\phi_{\rm in}/\phi_c,\chi_{\rm in}/\chi_0\}$. 
}
\label{2Dfield}
\end{figure}

\begin{figure}
\includegraphics[height=2.5in,width=2.5in]
{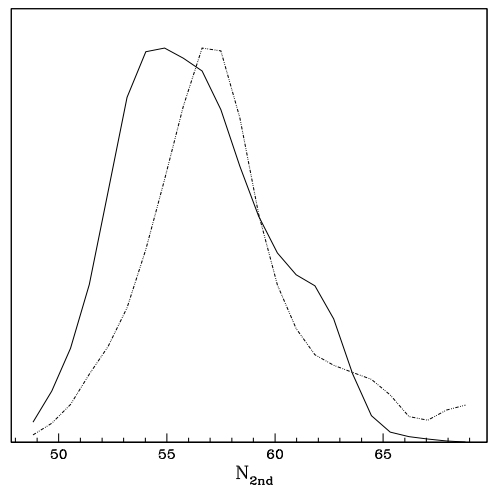}
\caption{Marginalised 1-d likelihood of the number 
of $e$-folds occurring during the 2nd phase of inflation.  The solid line
is based on the number of MCMC points, while the dotted
line weighs each point based on its $\bar{\chi}^2$ 
value, as described in the caption to Fig. 1.
}
\label{2ndefolds}
\end{figure}

In Fig.\,\ref{2Dfield} we plot the likelihood constraints for the initial 
values of the
scalar fields. These are also constrained to lie 
in a narrow region in the range
$1.1<\phi_{\rm in}/\phi_c<1.8$ and 
$2 \times 10^{-3}<\chi_{\rm in}/\chi_0<8 \times 10^{-3}$
(here $\chi_0 \equiv M/\sqrt{\lambda}$).
This means that initial values of $\phi$ close to $\phi=\phi_c$ are favoured.  
Since $m_\chi$ is much larger 
than $H$ for $\phi \gg \phi_c$,
the field $\chi$ is strongly suppressed for the initial conditions 
$\phi_{\rm in}/\phi_c \gg 1$. 
This corresponds to the case in which the perturbations 
exhibit violent growth in the tachyonic region, thus effectively ruled 
out in our linear analysis.
The initial value of $\chi$ affects the number of $e$-folds during the 
second stage of inflation ($=N_{\rm 2nd}$). 
We obtain smaller $N_{\rm 2nd}$ for larger $\chi_{\rm in}/\chi_0$.
As we find in Fig.\,\ref{2ndefolds}, the likely values for the number of  
$e$-folds is $50 \lesssim N_{\rm 2nd} \lesssim 65$
which corresponds to initial conditions 
$\chi_{\rm in}/\chi_0$ of order $10^{-3}$--$10^{-2}$.

It is rather surprising that the likelihood contours of $N_{\rm 2nd}$
are clustered in the region with cosmologically relevant scales.
In order to obtain this result we did not put any prior 
for the maximum values of the total number of $e$-folds.
We found that it is difficult to satisfy the conditions of 
COBE normalization and suppressed isocurvature 
perturbations unless $N_{\rm 2nd}$ ranges in the region  $50 \lesssim N_{\rm 2nd} \lesssim 65$.
This implies that double inflation has a rich and  complex 
structure relative to single-field inflation.

%%%%%%%%%%
%%%%%%%%%%

%%%%%%%%%%
%%%%%%%%%%

%%%%%%%%%%%%%%%%%%%%%%%%%%%%%%%%
\subsection{Power Spectrum}
%%%%%%%%%%%%%%%%%%%%%%%%%%%%%%%%

\begin{figure}
\includegraphics[height=2.5in,width=3.3in]
{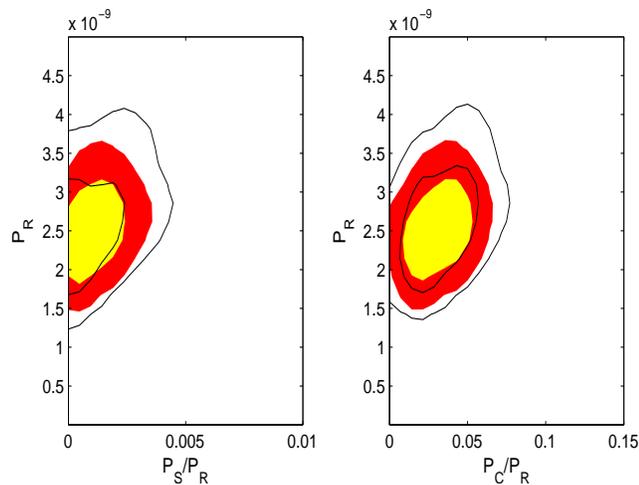}
\caption{2-dimensional likelihood contours for the amplitudes of
the power spectra, with adiabatic against isocurvature (left) and adiabatic
against correlated (right). Despite the freedom in
allowing running of the spectra index the isocurvature
component is severely constrained.
}
\label{2DPPS}
\end{figure}

In this subsection we consider the contribution of isocurvature 
perturbations to the CMB anisotropies.
In Fig.\,\ref{2DPPS} we plot observational contour 
bounds for the amplitude ${\mathcal{P}}_{\cal R}$ and 
the two ratios ${\mathcal{P}}_S/{\mathcal{P}}_{\cal R}$,
${\mathcal{P}}_C/{\mathcal{P}}_{\cal R}$.
The most likely value of ${\mathcal{P}}_{\cal R}$ 
is around ${\mathcal{P}}_{\cal R}=2.5 \times 10^{-9}$,
which is similar to the case of single-field 
inflation \cite{WM,single}.

\begin{figure}
\includegraphics[height=3.5in,width=3.5in]{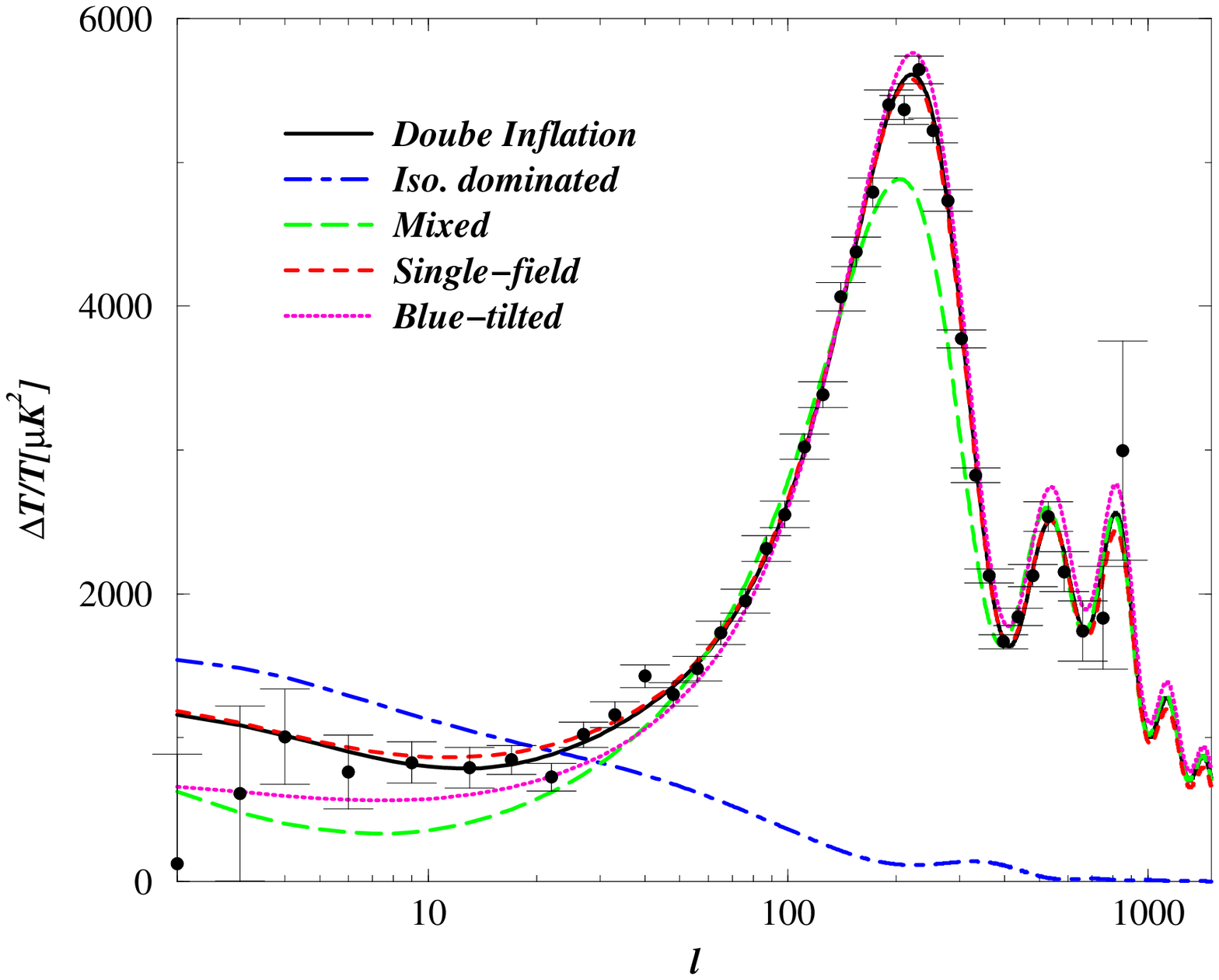}
\caption{The CMB angular power spectra for five different cases:\\
(i) our best-fit double inflation model,\\
(ii) isocurvature dominating over the adiabatic,\\\ (iii)
the isocurvature is comparable to the adiabatic (mixed),\\ 
(iv) the best fit single-field model with potential (\ref{posingle})
and \\ 
(v) a model with blue-tilted spectrum ($n_s>1$) on large scales. The spectra are significantly different 
from the standard one when the isocurvature is dominant.
}
\label{ttspec}
\end{figure}

The contribution of isocurvature perturbations is required 
to be small relative to adiabatic ones to be 
compatible with CMB anisotropies.
As shown in Fig.\,\ref{ttspec} the TT spectrum in the 
isocurvature dominated case does not fit with the WMAP
data at all. When isocurvature perturbations are comparable 
in magnitude to the adiabatic spectrum
(labelled ``mixed''), the spectrum shows significant  
deviations from 
the WMAP data on larger scales.
We found the $2\sigma$ bounds:
${\mathcal{P}}_S/{\mathcal{P}}_{\cal R}<0.004$
and ${\mathcal{P}}_C/{\mathcal{P}}_{\cal R}<0.07$
in order to be consistent with WMAP.

\begin{figure}
\includegraphics[height=3.0in,width=3.0in]{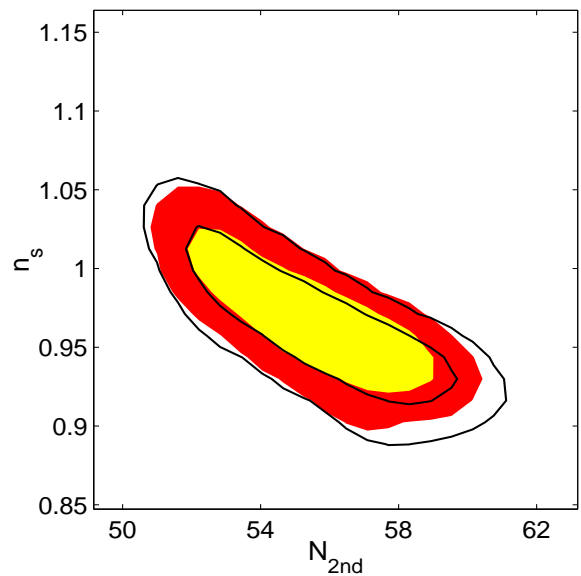}
\caption{2-dimensional likelihood contours for the spectral 
index $n_s$ and the number of $e$-folds during the 
second stage of inflation $N_{\rm 2nd}$.}
\label{nsN2}
\end{figure}

In Fig.\,\ref{nsN2} we plot the observational contour bounds on $N_{\rm 2nd}$ and the spectral index $n_s$.
There are some regions in which the spectrum of scalar perturbations 
is blue-tilted ($n_s>1$) with $N_{\rm 2nd} 
\lesssim 55$.
Since the power spectra generated in the 
first and second stages of inflation 
are blue- and red-tilted respectively \cite{Tsuji}, it is possible to have 
some suppression of power at low multipoles provided that 
the number of $e$-folds during the second stage of inflation satisfies $51 \lesssim N_{\rm 2nd} \lesssim 55$.
We show one example of the power spectrum in such a case 
in Fig.\,\ref{ttspec}. Although strong suppression around 
$\ell=2, 3$ is not easily achieved unless the spectrum is highly 
blue-tilted in this region (see e.g. \cite{Piao}), 
it is intriguing that this double inflation 
scenario provides a possibility to get a better 
fit on large scales.

%%%%%%%%%%
%%%%%%%%%%

%%%%%%%%%%
%%%%%%%%%%

%%%%%%%%%%
%%%%%%%%%%

%%%%%%%%%%%%%%%%%%%%%%%%%%%%%%%%
\section{Double inflation versus single-field inflation}
%%%%%%%%%%%%%%%%%%%%%%%%%%%%%%%%

A natural question is whether the extra complexity and fine-tuning 
involved in 
double inflation is actually preferred by the data over standard 
single-field inflation. This can be addressed by using the Akaike 
information criterion (AIC) 
and Bayesian Information criterion (BIC) \cite{bic}.  
These two criteria are defined as:
%%%%%%%%%
\begin{eqnarray}
{\rm AIC} & = & -2\ln\,{\cal L} + 2K \,,\\  
{\rm BIC} & = & -2\ln\,{\cal L} + K\ln \,N_p \,.
\end{eqnarray}
%%%%%%%%%
Here ${\cal L}$ is the maximum value of the likelihood, 
$K$ is the number of parameters 
and $N_p=1348$ is the number of WMAP data points.  
The optimal model minimises
the AIC or BIC. 
In the limit of large $N_p$, AIC tends to favour models with more
parameters while BIC more strongly penalises them (since the second term 
diverges in
this limit). BIC provides an estimate of the posterior evidence of a model 
assuming no prior information. Hence
BIC is a useful approximation to a full evidence calculation when we have 
no prior on the set of 
models (in this case single versus double inflation).
 In this case, we have no strong reason {\em a priori} to favour 
double inflation over single
field inflation so BIC provides sensible approximation to a full evidence calculation.

Our double inflation model has 5 inflationary parameters 
($M$, $m$, $\lambda$, $\phi_{\rm in}$, $\chi_{\rm in}$).  
We compare this with a single-field scenario with potential
%%%%%%%%%
\begin{eqnarray}
V=\frac{\lambda}{4}\left(\chi ^{2}
-\frac{M^{2}}{\lambda }\right)^{2}\,.
\label{posingle}
\end{eqnarray}
%%%%%%%%%
This has 3 inflationary parameters ($\lambda$, $M$, $\chi_{\rm in}$).
There are also 4 cosmological parameters, common to both models.
In Table \ref{criteriatable} we show the best-fit $\bar{\chi}^2$ and the 
values taken by the criteria for the models we have considered.

%%%%%%%%%
\begin{table}
\begin{tabular}{lccc}
\hline
Model & $-2\ln \,{\cal L}$ & AIC & BIC \\
\hline
Double inflation &  1428.85  & 1446.85 & 1493.70 \\ \\
Single-field & 1430.99   &  1444.99 & 1480.43 \\
\hline
\end{tabular}
\caption{\label{criteriatable}  
The best-fit $\bar{\chi}^2$ ($=-2\ln \,{\cal L}$) and best
Akaike and Bayesian Information 
criteria (AIC and BIC) for single and double inflation. Both criteria 
favour
single field inflation.}
\end{table}
%%%%%%%%%

We find that the best-fit value of $-2\ln \,{\cal L}$ in double inflation 
is smaller than in the case of single-field inflation.
However both the AIC and BIC values for double inflation are significantly
larger than those in the latter case, which suggests that single-field 
inflation is favoured relative to double inflation. 
In addition one could argue that single light-field
inflation should theoretically be preferred {\em a priori} since it does 
not require fine-tuning to achieve more than one field to be light relative to the Hubble constant. 
Adding this prior will
further favour single-field inflation.  

We have only included WMAP data. Evidence for running of the spectral 
index from WMAP and lyman-$\alpha$ data \cite{run} 
would favour double inflation models in which tilt 
is generic \cite{Tsuji}. However evidence for running is currently weak 
\cite{recentrun} 
and hence should not affect our conclusions significantly. Strong evidence 
for running in 
future data might change the situation however.

%%%%%%%%%%%%%%%%%%
\section{Conclusions}
%%%%%%%%%%%%%%%%%%

In this paper we have studied observational constraints on 
double inflation using the WMAP first year data.
The model we adopted is the supersymmetric hybrid potential 
given in Eq. (\ref{po}). The presence of a tachyonic instability 
region after symmetry breaking leads to the correlation 
between adiabatic and isocurvature perturbations, which can 
significantly alter the CMB power spectrum compared to 
the case of adiabatic perturbations alone.

Comparing with first year WMAP CMB data we found that the correlated 
isocurvature component can be at most $7\%$ of the 
total contribution which is dominated by the  
adiabatic spectrum.

We carried out likelihood analysis in terms of 5 inflationary 
parameters and 4 cosmological parameters.
The likelihood values of inflationary parameters are clustered 
in a narrow region around $M/m_p \sim 5.0 \times 10^{-8}$, 
$\lambda \sim (M/m_p)^2$ and $m/M \sim 0.5$
(see Fig.\,\ref{2Dinf}).

In spite of the large number of freedom of model parameters
relative to single-field inflation, the parameter space of
double inflation is severely constrained.
This comes from the fact that it is not so easy to satisfy 
all constraints including COBE normalization and 
sufficiently suppressed isocurvature perturbations.

We also found that the number of $e$-folds in the 
second stage of inflation are constrained to 
lie in the range $50 \lesssim N_{\rm 2nd} \lesssim 65$. 
Loss of power on large scales (relevant to 
achieving suppressed CMB low multipoles) is 
possible when the number of $e$-folds is 
around $51 \lesssim N_{\rm 2nd} \lesssim 55$. 

We also compared double inflation with single-field inflation
by using the Akaike (AIC) and Bayesian 
information criteria (BIC).
While the minimum value of $\bar{\chi}^2$ in double inflation is slightly smaller 
than in single-field inflation, the information criteria strongly support 
single-field inflation over 
the supersynmmetric hybrid double inflation models we
studied.

Nevertheless we need to caution that the minimum $\bar{\chi}^2$ is 
still larger than the number of data points $N_p$
in current observations. 
We expect that future high-precision data such as the Planck 
satellite will provide more sophisticated information
to distinguish between double inflation and 
single-field inflation. 

In this regard it will be interesting to extend 
our analysis to include more fields, so that 
the matter power spectrum can exhibit sharp features,
and to allow more realistic treatment of reheating. 
Both of these will increase the number of 
inflationary parameters (by about 2 or 3 each) and
it is difficult to imagine them producing 
smaller values of the AIC and BIC as a result.

It seems likely therefore that single field 
inflation will continue to be the scenario to beat.
It is intriguing that both the early and late universe seem well-described by very 
simple inflationary stages, and perhaps even two pure de Sitter phases.  Finding a theoretical basis for 
this perplexing high-energy/low-energy duality may 
become a dominant quest in cosmology 
in the coming years.

%%%%%%%%%%%%%%%%%%%%%%%%%%%%%
\section*{ACKNOWLEDGEMENTS}
We thank Nicola Bartolo, Pedro Ferreira and Andrew Liddle for useful 
discussions. S.T. is grateful to Rome observatory, 
Universities of Sussex, Queen Mary 
and Portsmouth for their warm hospitality 
during which
part of this work was done.  B.B. is supported by a Royal Society-JSPS 
fellowship.
The analysis was carried out on the 
multiprocessor machines Solent and Vela in Portsmouth and the UK national 
cosmology supercomputer (COSMOS) in Cambridge.
%%%%%%%%%%%%%%%%%%%%%%%%%%%%%

\end{document}